\documentclass[a4paper]{article}
\usepackage[english]{babel}
\usepackage[centertags]{amsmath}
\usepackage{latexsym, amsfonts, amssymb, amsthm}

\usepackage{fullpage}

\usepackage[all,dvips]{xy}
\usepackage{graphicx}
\usepackage[latin1]{inputenc}
\usepackage{url}
\usepackage{enumerate}
\usepackage[blocks]{authblk}

\theoremstyle{plain}
\newtheorem{thm}{Theorem}[section]
\newtheorem{lem}[thm]{Lemma}
\newtheorem{prop}[thm]{Proposition}
\newtheorem{cor}[thm]{Corollary}

\theoremstyle{definition}
\newtheorem{defn}[thm]{Definition}

\theoremstyle{remark}
\newtheorem{rem}[thm]{Remark}

\newtheorem{exmp}[thm]{Example}

\def\P{{\mathbf P}}

\def\H{\mathcal{H}}
\def\E{\mathcal{E}}
\def\A{\mathbf{A}}
\newcommand{\F}{\mathbf{F}}
\newcommand{\isomto}{\overset{\sim}{\longrightarrow}}
\newcommand{\pr}{\partial}

\title{Evaluation codes from smooth quadric surfaces and twisted Segre varieties}
\author{Alain Couvreur\thanks{Universit\'e Bordeaux I -- Institut de Math\'ematiques de Bordeaux} and Iwan Duursma\thanks{University of Illinois at Urbana-Champaign -- Department of Mathematics}}

\begin{document}

\maketitle

\begin{abstract}
We give the parameters of any evaluation code on a smooth quadric surface.
For hyperbolic quadrics the approach uses elementary results on product codes and the parameters of codes on elliptic quadrics are obtained by detecting a BCH structure on these codes and using the BCH bound.
The elliptic quadric is a twist of the surface $\P^1 \times \P^1$ and we detect a similar BCH structure on twists of the Segre embedding
of a product of any $d$ copies of the projective line.
\end{abstract}

\bigbreak


\bigbreak


\section*{Introduction}
The parameters of evaluation codes on quadric surfaces have been studied by Aubry (who also considered higher dimensional quadrics) in \cite{aubry} and by Edoukou in \cite{fred2}.
Most of the results on the topic concern the evaluation of forms of degree one or two.
The reason of this restriction is that the estimate of the minimum distance of such codes by geometric methods becomes harder when the degree increases.

In this article, we give the parameters of all evaluation codes on smooth quadric surfaces. The approach is not based on point counting but on the detection of a particular structure on the codes.
Namely, we prove that
codes on hyperbolic quadrics are tensor products of two extended Reed--Solomon codes and that codes on elliptic quadrics are extensions of some BCH codes studied by Pellikaan and the second author in \cite{DuursmaPellikaan}.
A nice consequence of these results is that they solve a point counting problem which was not proved up to now.
It should be underlined that usually, one tries to estimate the parameters of an Algebraic Geometry code by solving some equivalent geometric problem.
In the present paper we proceed in the opposite direction.
Namely, we are able to solve open geometric problems using known coding theoretic results.

Basically, studying codes on hyperbolic and elliptic quadrics reduces to study codes on $\P^1\times \P^1$ and a twist of it.
This approach has a natural generalisation to products of $d\geq 2$ copies of $\P^1$ yielding naturally tensor products of $d$ extended Reed--Solomon codes and their twists yielding extended BCH codes of length $q^d+1$.
In particular, this construction gives a geometric realisation of a large class of BCH codes as evaluation codes and without using a subfield subcode operation.

The paper is organised as follows. The prerequisites on evaluation codes, twists and quadric surfaces are given in Section \ref{SecPrerequisites}. Evaluation codes on hyperbolic quadric surfaces are considered in Section \ref{SecHyper} and codes on elliptic quadrics are treated in Section \ref{SecEll}. The higher dimensional case is studied in Section \ref{SecHigher}.

\section{Prerequisites}\label{SecPrerequisites}

\subsection{Evaluation codes}
Consider the projective space $\P^r_{\F_q}$ with its coordinate ring $\F_q [x_0, \ldots , x_r]$.
For an integer $s$, denote by $\mathcal{F}_r(s)$ the space of homogeneous forms of degree $s$ in $r+1$ variables, i.e. the space $H^0 (\P^r, \mathcal{O}_{\P^r}(s))$.
Given $f\in \mathcal{F}_{r}(s)$ and $P$ a point of $\P^r$, we define the \emph{evaluation} of $f$ at $P$ as $f(P):=f(p_0, \ldots , p_r)$, where $(p_0:\ldots :p_r)$ is the system of homogeneous coordinates of $P$ such that the first nonzero coordinate starting from the left is set to $1$, i.e. is of the form $(0:\ldots:0:1:p_i:\ldots :p_r)$.

\begin{defn}
Let $X\subset \P^r$ be a smooth projective variety over $\F_q$.
The evaluation code $C_X(s)$ is defined as the image of the evaluation map
$$
ev:\left\{
  \begin{array}{ccc}
    \mathcal{F}_{r}(s) & \longrightarrow & \F_q^n\\
    f & \longmapsto & (f(P_1), \ldots , f(P_n))
  \end{array}
\right. ,
$$
where $P_1, \ldots , P_n$ are the $\F_q$--points of $X$.  
\end{defn}

\noindent If we denote by $I_X (s)$ the degree $s$ part of the homogeneous ideal $I_X\subset \F_q [x_0,\ldots , x_r]$ associated to $X$, then the above map $ev$ obviously factors as $ev: \mathcal{F}_r(s)/I_X(s)\longrightarrow \F_q^n$.

The codes $C_X(s)$ for $X=\P^r$ are the projective Reed-Muller codes $PC_s(r,q)$ whose parameters were obtained by S{\o}rensen \cite[Theorem 1]{Sor91}.
In this paper, we first consider the case that $X \subset \P^3$ is a smooth quadric. The case of a hyperbolic quadric corresponds to the Segre embedding of 
$\P^1 \times \P^1$ in $\P^3$ and the case of an elliptic quadric to a twist of such an embedding. We will then consider more generally the case that $X$ is the 
Segre embedding of the product $\P^1 \times \cdots \times \P^1 \hookrightarrow \P^{2^d-1}$ of $d$ copies of the projective line, or a twist of such an embedding. 



\subsection{Twists}
 Given two varieties $X$ and $Y$ over a field $k$, one says that $Y$ is a twist of $X$ if the two varieties are not isomorphic as $k$--varieties but are as $K$--varieties, where $K$ is a finite extension of $k$.
For instance, the plane curves over $\mathbf{Q}$ defined by the homogeneous equations $x^2+y^2-z^2=0$ and $x^2+y^2+z^2=0$ are $\mathbf{Q}(\sqrt{-1})$--isomorphic but not $\mathbf{Q}$--isomorphic.

\subsection{Smooth quadric surfaces}

\subsubsection{Elliptic and hyperbolic quadrics}

Over a finite field $\F_q$ there exist two distinct isomorphism classes of smooth quadric surfaces, respectively called \emph{elliptic quadrics} and \emph{hyperbolic quadrics}.
In $\P^3$, a hyperbolic quadric is projectively equivalent to the surface of equation
\begin{equation}
  \label{EqHyper}
  x_0 x_3-x_1 x_2 =0.
\end{equation}
Given an irreducible homogeneous polynomial $Q(x,y)$ of degree two over $\F_q$, then any elliptic quadric is projectively equivalent to the surface of equation
\begin{equation}
  \label{EqEll}
  x_0 x_3 - Q(x_1, x_2) = 0.
\end{equation}
We refer the reader to \cite{Hirschfeld} for further details on these surfaces.

\begin{rem}\label{TwistingRem}
One can easily prove that the elliptic quadric is a twist of the hyperbolic one. Let $(x+wy)(x+w^qy)$ be the factorisation of $Q$ over $\F_{q^2}$ (with $w\in \F_{q^2}\setminus \F_q$). The $\F_{q^2}$--linear automorphism of $\P^3$
\begin{equation}\label{TwistingMap}
\mu_{tw} : P \longmapsto AP,\qquad 
A = \left( \begin{array}{llll} 1 &0 &0 &0 \\ 0 &1 &\omega &0 \\ 0 &1 &\omega^q &0 \\ 0 &0 &0 &1 \end{array} \right).
\end{equation}
 induces an $\F_{q^2}$--isomorphism between the elliptic and the hyperbolic quadric. 
\end{rem}

\subsubsection{Rational parametrisation of quadrics.}
Elliptic and hyperbolic quadrics are both rational. Here is a birational map from $\P^2$ to the hyperbolic quadric defined in (\ref{EqHyper}).
\begin{equation}\label{EmbeddingHyper}
\left\{\begin{array}{ccc}
  \P^2 & \dashrightarrow & \P^3 \\
  (x:y:z) & \longmapsto & (z^2:xz:yz:xy)
\end{array}\right. .
\end{equation}

\noindent Here is a birational map from $\P^2$ to the elliptic quadric defined in (\ref{EqEll}).
\begin{equation}\label{EmbeddingEll}
\left\{\begin{array}{ccc}
  \P^2 & \dashrightarrow & \P^3 \\
  (x:y:z) & \longmapsto & (z^2:xz:yz:Q(x,y))
\end{array}\right. .
\end{equation}

\begin{rem}\label{DefinitionEmbeddingHyper}
 The map (\ref{EmbeddingHyper}) is regular on $\P^2\setminus \{(1:0:0), (0:1:0)\}$. Denote by $C$ the subvariety $\mathcal{H}\cap \{x_0=0\}$, then the image of the map (\ref{EmbeddingHyper}) is $(\mathcal{H} \setminus C) \cup \{(0:0:0:1)\}$.  
\end{rem}

\begin{rem}\label{DefinitionEmbeddingEll}
  Let $P$ be the closed point of degree $2$ of $\P^2$ defined by
$$\{P\}=\{z=0\}\cap \{Q(x,y)=0\},$$
then the map (\ref{EmbeddingEll}) is regular on $\P^2\setminus \{P\}$. It is in particular regular at all the $\F_q$--rational points of $\P^2$. Denote by $C$ the subvariety $\mathcal{E}\cap \{x_0=0\}$, then the image of the map (\ref{EmbeddingHyper}) is $(\mathcal{E} \setminus C) \cup \{(0:0:0:1)\}$.  
It is worth noting that the unique $\F_q$--rational point of $C$ is $(0:0:0:1)$. Thus, the image of the map (\ref{EmbeddingEll}) contains all the rational points of $\E$.

\end{rem}

\section{Codes from hyperbolic quadrics}\label{SecHyper}

From now on, the hyperbolic quadric is denoted by $\mathcal{H}$.
It is well--known that $\H$ is isomorphic to $\P^1\times \P^1$.
Indeed, the quadric $\H$ with equation $x_0 x_3-x_1 x_2=0$ 
is the image of the Segre embedding (see \cite[Chapter 4 \S 4]{fulton}, \cite[Chapter I \S 5.1]{sch1}):
\begin{equation}\label{Segre}
\phi_{s} : \left\{
  \begin{array}{ccc}
    \P^1 \times \P^1 & \longrightarrow & \P^3\\ 
    ((u_0:v_0),(u_1:v_1)) & \longmapsto & (u_0 u_1 : u_0 v_1 : v_0 u_1 : v_0 v_1)
  \end{array}
\right. .
\end{equation}
A homogeneous form $f\in \mathcal{F}_3 (s)$ pulled back by $\phi$ yields the bi-homogeneous form $f(u_0 u_1, u_0 v_1, v_0 u_1, v_0 v_1)$ of bi-degree $(s,s)$.
Afterwards, one sees easily that the pullback map $\phi^{\star}$ induces an isomorphism
$\mathcal{F}_3(s)/I_{\H}(s)\isomto \mathcal{F}_1(s) \otimes \mathcal{F}_1(s)$,
where $I_{\H}(s)$ is the degree $s$ part of the homogeneous ideal associated to $\H$.
Consequently, the code $C_{\H}(s)$ is nothing but the code $C_{\P^1}(s)\otimes C_{\P^1}(s)$. The code $C_{\P^1}(s)$ is an extended Reed--Solomon code with parameters $[(q+1),(s+1),q-s+1]$. It is well--known that the minimum distance of a tensor product of two codes is the product of the minimum distances. This yields the following result. 
  
\begin{thm}\label{CodeHyper} 
  Let $\H$ be a hyperbolic quadric over $\F_q$, let $s$ be an integer such that $s< q$, then the code $C_{\H}(s)$ has parameters $[(q+1)^2, (s+1)^2, (q-s+1)^2]$.
\end{thm}

\begin{rem}\label{hansen}
  The above result is already partially proved by S.H. Hansen in 
 \cite[Example 3.2]{Han01}, where the author obtains $(q-s+1)^2$ as a lower bound for the minimum distance without proving that it is reached.

Actually, Hansen considers more general evaluation codes on $\H$: the codes obtained by evaluating spaces of forms whose pullback by $\phi$ are of the form $\mathcal{F}_1(a)\otimes\mathcal{F}_1(b)$.
Using the above approach, one proves easily that such codes have parameters $[(q+1)^2, (a+1)(b+1), (q-a+1)(q-b+1)]$. This proves that the lower bound of Hansen is the actual minimum distance.
\end{rem}

\begin{rem}
  For $s=2$, the result has been proved in \cite[Theorem 6.2]{fred2}.
\end{rem}

\begin{rem}\label{allcodes}
  Using the structure of the Picard group of $\H$, one can prove that any evaluation code on $\H$ is equivalent to one of the codes described in Remark \ref{hansen}. Therefore, using Remark \ref{hansen}, we have the exact parameters of any evaluation code on $\H$.
\end{rem}

Theorem \ref{CodeHyper} has the following geometric corollary.

\begin{cor}[Maximum number of points of an $(s,s)$--curve]\label{cor:Mini_hyper}
Let $X$ be a curve obtained by the intersection of $\H$ with a surface of degree $s$ of $\P^3$ which does not contain $\H$. Then, the number of rational points of $X$ satisfies
$$
\sharp X(\F_q)\leq 2s(q+1)-s^2
$$
and the equality holds if and only if $X$ is a union of $s$ lines of the form $\phi(\{a\}\times \P^1)$ and $s$ lines of the form $\phi(\P^1 \times \{b\})$.
\end{cor}

\begin{proof}
  The upper bound comes from Theorem \ref{CodeHyper}.
Moreover, it is easy to see that the union of $s$ rational lines of the first ruling and $s$ lines of the other one has $2s(q+1)-s^2$ rational points.

Conversely, it is well--known that the minimum weight codewords of a tensor product of codes are tensor products of minimum weight codewords. Thus, minimum weight codewords of $C_{\H}(s)$ are obtained by the evaluation of forms $f$ whose pullback $\phi^{\star}f$ equals $g(u_0,v_0)h(u_1,v_1)$, where $g,h$ both split in products of $s$ distinct polynomials of degree one.
Thus, the vanishing locus of $f$ is a union of lines and any $f$ whose vanishing locus is not such a union has strictly less rational points in its vanishing locus on $\H$. 
\end{proof}

\subsubsection*{About the geometry of the minimum weight codewords of $C_\H (s)$}

  In Corollary \ref{cor:Mini_hyper}, one can prove easily that if $\sharp X (\F_q)=2s(q+1)-s^2$, then, one of the surfaces $\mathcal{S}$ of degree $s$ such that $\mathcal{S}\cap \H=X$ is a union of $s$ distinct planes such that each one of them is tangent to $\H$ at some rational point. This claim generalises \cite[Theorem 6.3]{fred2}, which treats the case $s=2$.

  \begin{rem}\label{rem:Thm_Edoukou}
    In \cite[Theorem 6.3]{fred2}, the author asserts that if $X=\H \cap \mathcal{S}$, with $\mathcal{S}$ a quadric, has $4q$ rational points, then $\mathcal{S}$ is either a pair of planes or another hyperbolic quadric with $4$ common lines with $\H$.
Actually, the set of quadrics $\mathcal{S}$ such that $\mathcal{S}\cap \H=X$ is a linear system of dimension $1$ which always contains a pair of planes.    
  \end{rem}

\section{BCH codes and codes on elliptic quadrics}\label{SecEll}

From now on, the elliptic quadric is denoted by $\E$ and $s$ denotes a positive integer.
The aim of this section is to prove that the codes $C_\E(s)$ are extended BCH codes. More precisely, these codes of length $q^2+1$ (the elliptic quadric has $q^2+1$ rational points, see \cite[Table IV.15.4]{Hirschfeld}) punctured at two positions yield BCH codes of length $q^2-1$. 
 
\subsection{The cyclic structure}
The cyclic structure of the punctured codes can be explained geometrically. Indeed, the automorphism group of the elliptic quadric contains an element fixing two rational points and shifting cyclically the $q^2-1$ other ones.

Let us describe such an automorphism. Consider the description of $\E$ given in (\ref{EqEll}) and assume moreover that $w$ is a primitive element of $\F_{q^2}/\F_q$. The multiplication by $w$ in $\F_{q^2}$ provides an automorphism $\sigma_w \in \mathbf{Aut}_{\F_q}(\A^2)$ which extends to $\mathbf{Aut}_{\F_q}(\P^2)$ and, thanks to the parametrisation map (\ref{EmbeddingEll}), yields an automorphism $\tilde{\sigma}_w\in \mathbf{Aut}_{\F_q}(\E)$. The map $\tilde{\sigma}_w$ is the restriction to $\E$ of a linear automorphism of $\P^3$ described by the matrix
$$
\left(
  \begin{array}{cccc}
    1 & 0 & 0 & 0 \\
    0 & 0 & -N(w) & 0 \\
    0 & 1 & Tr(w) & 0 \\
    0 & 0 & 0 & N(w)
  \end{array}
\right),
$$
where $N(w)$ and $Tr(w)$ denote respectively the norm $N(w):=w^{q+1}$ and the trace $Tr(w):=w+w^q$.
One can check that this automorphism fixes the points $(1:0:0:0)$ and $(0:0:0:1)$ and shifts cyclically the $q^2-1$ other rational points of $\E$.

\subsection{A class of BCH codes}\label{SubSectionBCH}

\subsubsection{The cyclic codes}

\begin{defn}\label{DefBCH}
For a given field $\F_q$ and a positive integer $s<q$, let $B(s)$ be the cyclic code defined over the extension 
field $\F_{q^2}$ which is generated by the vectors of the form $(\zeta^r | \zeta \in \F_{q^2}^{\times})$,
for $r=i+qj$ such that $0 \leq i,j\leq s$.
In addition, let $B_0(s)$ be the subfield subcode ${B(s)}_{|\F_q}$. 
\end{defn}

This class of codes is studied in \cite{DuursmaPellikaan} where the following result is proved.  

\begin{prop}\label{DuursPell}
  The code $B_0(s)$ has parameters $[q^2-1,(s+1)^2, q^2-1-s(q+1)]$. Moreover it is a BCH code.
\end{prop}

\begin{proof}
  \cite[Proposition 12]{DuursmaPellikaan}. \end{proof}

\begin{rem}\label{subfield}
The condition $0 \leq i, j \leq s$ differs from the condition
$0 \leq i+j \leq s$ that is used to describe punctured Reed-Muller codes as cyclic codes.  
\end{rem}

\subsubsection{The extended BCH codes}
Actually the codes from elliptic quadrics are related to some extended version of the above described BCH codes. Thus, we introduce a new class of codes.

\begin{defn}\label{DefExtBCH}
Consider the projective line $\P^1$ over $\F_{q^2}$ and let $0\leq s\leq q-1$ be an integer. We denote by $B^{ext}(s)$ the subcode of $C_{\P^1}(s(q+1))$ spanned by the evaluation at $\P^1(\F_{q^2})$ of the forms
$$
x^{i+qj}y^{s-i+q(s-j)},\qquad 0\leq i,j \leq s.
$$ 
The extended BCH code $B_0^{ext}(s)$ is defined as the subfield subcode $B^{ext}(s)_{|\F_q}$.
\end{defn}

\begin{rem}\label{Punct}
  Clearly, $B (s)$ and $B_0(s)$ can be respectively obtained by puncturing $B^{ext}(s)$ and $B_0^{ext}(s)$ at the positions corresponding to $(0:1)$ and $(1:0)$.
\end{rem}

\begin{rem}\label{BigGroup}
  An interesting feature of the codes $B_0^{ext}(s)$ compared to $B_0(s)$ is that they have a large permutation group. Indeed, the group $PSL(2, \F_{q^2})$ acts on $B^{ext}(s)$ and  $B_0^{ext}(s)$ by permutation. In particular, these codes are $3$--transitive.
\end{rem}

\begin{prop}\label{ParamExtBCH}
  For $0\leq s \leq q-2$, the code $B_0^{ext}(s)$ has parameters $[q^2+1, (s+1)^2, q^2+1-s(q+1)]$.
\end{prop}

\begin{proof}
  The length is obvious. For the dimension, let us prove that the puncturing map $p:B_0^{ext}(s) \longrightarrow B_0(s)$ evoked in Remark \ref{Punct} is injective. Denote respectively by $P_0$ and $P_{\infty}$ the points $(0:1)$ and $(1:0)$ of $\P^1$. The kernel of $p$ is the subspace of codewords of $B_0^{ext} (s)$ with supports contained in $\{P_0, P_{\infty}\}$. If such a nonzero word exists, then from Remark \ref{BigGroup}, there exists a word of weight $\leq 2$ whose support avoids $P_0$ and $P_{\infty}$. By puncturing, this would yield a codeword of weight $\leq 2$ in $B_0(s)$, which contradicts Proposition \ref{DuursPell}.

For the minimum distance, using Proposition \ref{DuursPell} and Remark \ref{Punct} we know that the minimum distance $d$ of $B_0^{ext}(s)$ satisfies
\begin{equation}\label{DistExt}
 d \leq q^2+1 -s(q+1).
\end{equation}
Take a codeword $w\in B_0^{ext} (s)$ of minimum weight $d$. Using Remark \ref{BigGroup}, one can assume that $P_0$ and $P_{\infty}$ are contained in the support of $w$. The punctured codeword $p(w)\in B_0(s)$ has weight $d-2$ and from Proposition \ref{DuursPell}, we have $d-2 \geq q^2-1-s(q+1)$. This inequality together with (\ref{DistExt}) yield the result.
\end{proof}

\subsection{A twisted embedding of the projective line}\label{TwistSubSec}

The elliptic quadric $\E \subset \P^3$ over $\F_q$ contains $q^2+1$ rational points. Using (\ref{EmbeddingEll}) together with Remark \ref{DefinitionEmbeddingEll} they are described by
\begin{equation}\label{SetP}
{\cal P} = \{ (1:u:v:Q(u,v)) : u,v \in \F_q \} \cup \{ (0:0:0:1) \},
\end{equation}
where $Q(u,v) = (u+\omega v)(u+\omega^q v)$ as in (\ref{EmbeddingEll}).

Set
\begin{equation}\label{embq0}
 \phi_f  :  \left\{\begin{array}{ccc}
  \P^1  & \longrightarrow & \P^1 \times \P^1 \\
  (x:y) & \longmapsto     & ((x:y),(x^q:y^q)) 
\end{array}\right. .  
\end{equation}


\noindent Consider the $\F_q$--embedding
\begin{equation} \label{embq1}
\psi : \left\{\begin{array}{ccccc}
  \P^1  & \overset{\phi_f}{\longrightarrow} & \P^1 \times \P^1 & \overset{\phi_s}{\longrightarrow} & \P^3 \\
  (x:y) & \longmapsto     & ((x:y),(x^q:y^q)) & \longmapsto  & (y^{q+1}:xy^q:x^qy:x^{q+1}) 
\end{array}\right. ,
\end{equation}

\noindent where $\phi_f$ is defined in (\ref{embq0}) and $\phi_s$ is the Segre embedding (\ref{Segre}).
Over $\F_{q^2}$, the projective line has $q^2+1$ rational points $\{ (x:1) : x \in \F_{q^2} \} \cup (1:0).$ Writing $x=u+\omega v$,
for $u,v \in \F_q$, their images in $\P^3$ by the map (\ref{embq1}) are
\begin{equation}\label{SetPprime}
{\cal P}' = \{ (1:u+\omega v: u+\omega^q v :Q(u,v)) : u,v \in \F_q \} \cup \{ (0:0:0:1) \}.
\end{equation}
Clearly, a point $P' \in {\cal P'}$ differs from a point $P \in {\cal P}$ by the linear transformation $\mu_{tw}$ of (\ref{TwistingMap}).
Consequently, we state the following lemma.

\begin{lem}\label{TheGoodMap}
We have an $\F_{q^2}$--embedding of $\P^1$ 
\begin{equation}\label{Fq2emb}
\psi_{tw} : = \mu_{tw}^{-1} \circ \psi: \P^1 \overset{\phi_f}{\longrightarrow} \P^1\times \P^1 \overset{\phi_s}{\longrightarrow} \P^3 \overset{\mu_{tw}^{-1}}{\longrightarrow} \P^3
\end{equation}
 inducing a one-to-one map $\P^1(\F_{q^2}) \longrightarrow \E (\F_q)$.  
\end{lem}

Thanks to the $\F_{q^2}$--embedding $\psi_{tw}$, the $\F_{q^2}$--codes $C_{\E}(s)\otimes \F_{q^2}$ can be regarded as codes over $\P^1$. This is the key point of the proof of the equality $C_{\E}(s)=B_0^{ext}(s)$ (up to a permutation) established in the following subsection.


\subsection{The parameters of the codes on the elliptic quadric}

The objective is to determine the parameters and in particular the minimum distance of the codes $C_\E(s)$. This objective is reached by Theorem \ref{Main}.
Recall that except for the case $s=1,2$, the minimum distance of these codes was unknown up to now.

For the proof of Theorem \ref{Main} we need the following combinatorial lemma.

\begin{lem} \label{lemUV}
The sets of pairs of integers $U^{(s)} = \{(i+k, j+k) : 0 \leq i, j, k \text{ and } i+j+k \leq s \}$ and $V^{(s)} = \{ (i,j) : 0 \leq i, j \leq s \}$ are equal.
\end{lem}

\begin{proof}
Clearly $U^{(s)} \subset V^{(s)}$. Conversely, for $(i,j) \in V^{(s)}$ and for $k = \min\{i,j\}$, we have $(i,j)=((i-k)+k, (j-k)+k) \in U^{(s)}.$
\end{proof}


\begin{thm}\label{Main}
  The code $C_\E (s)$ is permutation equivalent to the extended BCH code $B_0^{ext}(s)$ introduced in Definition \ref{DefBCH}. 
  Therefore, for all $0\leq s< q-1$, the code $C_{\E}(s)$ has parameters $[q^2+1, (s+1)^2, q^2+1-s(q+1)]$.
\end{thm}

\begin{proof}
  The code $C_\E(s)$, which is defined over $\F_q$, and the code $C_\E(s) \otimes \F_{q^2}$, which has coefficients over $\F_{q^2}$, use the same generator matrix and have the same parameters.

Clearly, the subfield subcode ${(C_{\E}(s)\otimes \F_{q^2})}_{|\F_q}$ equals $C_{\E}(s)$. Thus, to prove that $C_{\E}(s) = B_0^{ext}(s)$, it is sufficient to prove that $C_{\E}(s)\otimes \F_{q^2}=B^{ext}(s)$ (see Definition \ref{DefExtBCH}).
Afterwards, the parameters of $C_{\E}(s)$ are given by Proposition \ref{ParamExtBCH}.

\medbreak

\noindent {\it Step 1.} We first prove that $C_{\E}(1)\otimes \F_{q^2}=B^{ext}(1)$.
The code $C_{\E}(1)\otimes \F_{q^2}$ is obtained by evaluating $\mathcal{F}_3(1)\otimes \F_{q^2}$ at the set $\mathcal{P}$ described in (\ref{SetP}). 

Because of the bijection induced by $\psi_{tw}$ in Lemma \ref{TheGoodMap} between
the $\F_q$--rational points of $\E$ and the $\F_{q^2}$--rational points of $\P^1$, the code can equivalently be obtained by evaluating the pullbacks $\psi_{tw}^{\star} (\mathcal{F}_3(1)\otimes \F_{q^2})$ at the elements of $\P^1 (\F_{q^2})$.

Recall that, from \S \ref{TwistSubSec}, we have $\psi_{tw} = \mu_{tw}^{-1}\circ \phi$, where $\mu_{tw}$ and $\psi$ are respectively defined in (\ref{TwistingMap}) and (\ref{embq1}).
Since $\mu_{tw}$ is $\F_{q^2}$--linear, one sees easily that ${(\mu_{tw}^{-1})}^{\star} (\mathcal{F}_3(1)\otimes \F_{q^2}) = \mathcal{F}_3(1)\otimes \F_{q^2}$ and hence $\psi_{tw}^{\star} (\mathcal{F}_3(1)\otimes \F_{q^2}) = \psi^{\star} (\mathcal{F}_3(1)\otimes \F_{q^2}$).
Finally, (\ref{embq1}) entails that  $\psi^{\star} (\mathcal{F}_3(1)\otimes \F_{q^2})$ is generated by $y^{q+1}, xy^q, x^q y, x^{q+1}$. Evaluating these forms at $\P^1 (\F_{q^2})$ yields $B^{ext}(1)$ (see Definition \ref{DefExtBCH}).

\medbreak 

\noindent {\it Step 2.} For the general case we just copy Step 1. By the same manner $C_{\E}(s)\otimes \F_{q^2}$ can be obtained by evaluating the elements of $\psi_{tw}^{\star} (\mathcal{F}_3(s)\otimes \F_{q^2})$ at the elements of $\P^1 (\F_q)$. In addition, one proves, as in Step 1, that $\psi_{tw}^{\star} (\mathcal{F}_3(s)\otimes \F_{q^2}) = \psi^{\star} (\mathcal{F}_3(s)\otimes \F_{q^2})$.

The space $\psi^{\star} (\mathcal{F}_3(s)\otimes \F_{q^2})$ is generated by monomials of degree $s$ in $y^{q+1}, xy^q,$ $x^qy, x^{q+1}$. Such a monomial is of the form
$$
y^{a(q+1)} {(xy^q)}^b {(x^q y)}^c x^{d(q+1)} = x^{(b+d)+q(c+d)} y^{(a+c)+q(a+b)},\quad \textrm{for}\ a+b+c+d=s.
$$
From Lemma \ref{lemUV}, this set of monomials equals 
$$
\left\{x^{(i+qj)} y^{(s-i)+q(s-j)}\ |\ 0\leq i,j \leq s \right\},
$$
which yields the result by definition of $B^{ext} (s)$.
\end{proof}
 
\begin{rem}
  It is worth noting that the above proof points out a very interesting property of $B(s)$.
Indeed, even if $B(s)$ is defined over $\F_{q^2}$, it is generated by words defined over $\F_q$. Thus, the $\F_q$--dimension of its subfield subcode $B_0(s)$ equals the $\F_{q^2}$--dimension of $B(s)$. 
This explains why the codes $B_0(s)$ provide many of the best known codes (see \cite{grassl}): in general the subfield subcode operation entails a dramatic reduction of the dimension. This reduction does not happen for the codes $B(s)$.
\end{rem}

\begin{rem}
  Since the Picard group of $\E$ is generated by $\mathcal{O}_{\E}(1)$, any evaluation code on this surface is equivalent to $C_{\E}(s)$ for some $s$.
Thus, as for the hyperbolic quadric, we have here the exact parameters of any evaluation code on $\E$.
\end{rem}

Theorem \ref{Main} has a geometric corollary.

\begin{cor}\label{cor:mini_ell}
Let $s<q-1$.
  Let $X \subset \mathcal{E}$ be a curve obtained by the intersection of $\E$ with a surface of degree $s$ which does not contain $\E$.
Then,
$$
\sharp X(\F_q)\leq s(q+1).
$$ 
\end{cor}

\begin{proof} It is a straightforward consequence of Theorem \ref{Main}. \end{proof}

\subsubsection*{About the geometry of the minimum weight codewords of $C_{\E}(s)$}

Comparing Corollary \ref{cor:mini_ell} with Corollary \ref{cor:Mini_hyper}, it is natural to ask:  
{\it If equality holds in Corollary \ref{cor:mini_ell}, is the curve $X$ a cut out of $\E$ by $s$ planes?} 

\medbreak

Consider $s$ distinct planes $\Pi_1, \ldots ,\Pi_s$ non tangent to $\E$ and such that for all $i, j$, the line $\Pi_i\cap \Pi_j$ does not meet $\E$ at rational points and set $\mathcal{S}:= \Pi_1 \cup \ldots \cup \Pi_s$. Clearly, the curve $X:=\mathcal{S}\cap \E$ has $s(q+1)$ rational points. 
Conversely, if $s=1, 2$, the curves reaching this upper bound are always cut outs by $s$ planes. The claim is elementary for $s=1$ and the case $s=2$ is treated in \cite[Theorem 6.9]{fred2} (an argument similar to that of Remark \ref{rem:Thm_Edoukou} leads to this conclusion).
However, for $s\geq 3$, there exist curves reaching this bound but which are not cut outs by planes, some of them are actually irreducible. Computer aided calculations using the software {\sc Magma} \cite{magma} provided the following example.

\begin{exmp}
  Let $s=3$ and $q=5$. The surface $\E$ is defined by the equation $3y^2 + 3yz + z^2 + 4xt=0$.
Let $\mathcal{S}$ be the surface of equation 
$$
(\mathcal{S}) \qquad 3x^3 + 2x^2y + 2xy^2 + 3x^2z + 4xyz + 3y^2z + 2x^2t + 2xyt + 4xzt + 4yzt + xt^2 + 3yt^2 + 2zt^2=0,
$$
then the curve $X=\E \cap \mathcal{S}$ is irreducible and has $18= 3(5+1)$ rational points.
\end{exmp}

\section{Higher dimensional analogues}\label{SecHigher}

The results in the previous sections give us the actual parameters of evaluation codes on smooth quadric surfaces.
The case of a hyperbolic quadric was proved by establishing a relation with tensored Reed-Solomon codes and the
case of an elliptic quadric was proved using a correspondence with a suitable class of BCH codes. The hyperbolic quadric 
is the image $\H$ of $\P^1 \times \P^1$ in $\P^3$ under the Segre embedding and the elliptic quadric $\E$ is a
quadratic twist of this embedding. Both embeddings generalise and in this section we will describe evaluation codes 
defined on the image $\H \subset \P^{2^d-1}$ 
of the Segre embedding $\phi : \P^1 \times \cdots \times \P^1 \longrightarrow \P^{2^d-1}$ of $d$ copies of $\P^1$
and on twists $\E$ of $\H$. 

\subsection{The non-twisted case}

Let $\H \subset \P^{2^d-1}$ be the Segre embedding of $d$ copies of $\P^1$ and let $I_\H \subset \F_r$
be its associated homogeneous ideal. As in (\ref{Segre}) denote by $\phi_s$ the Segre's embedding. Similar to the case of the hyperbolic quadric, the pullback map $\phi_s^{\star}$ induces an isomorphism
$\mathcal{F}_d(s)/I_{\H}(s)\isomto \mathcal{F}_1(s) \otimes \cdots \otimes \mathcal{F}_1(s)$ ($d$ copies). Consequently, the evaluation code 
$C_{\H}(s)$ over $\F_q$ with $s < q$ can be described as a tensor product $C_{\P^1}(s)\otimes C_{\P^1}(s)$ of extended Reed-Solomon codes.

\begin{thm}\label{CodeHyper2} 
  Let $\H$ be the Segre embedding of the product $\P^1 \times \cdots \times \P^1 \hookrightarrow \P^{2^d-1}$ of $d$ copies of projective line over $\F_q$, let $s$ be an integer such that $s< q$, 
  then the code $C_\H(s)$ has parameters $[(q+1)^d, (s+1)^d, (q-s+1)^d]$. Moreover, the code is the
  $d$-fold tensor product of an extended Reed-Solomon code.
\end{thm}

It is well--known that the homogeneous ideal $I_\H \subset \mathcal{F}_{2^d-1} = \F_q [x_0, \ldots , x_r]$ for $\H$ is 
generated by quadrics. In fact this is true more generally for the larger class of Segre embeddings
of projective space of any dimension (details and further references can be found in \cite{Rub07}).
The Segre embedding $\H$ of $\P^1 \times \P^1 \times \P^1$ in $\P^7$ is the intersection of nine quadrics.
Here is a birational map from $\P^3$ to $\H$. 
\begin{equation}\label{EmbeddingHyper2}
\left\{\begin{array}{ccc}
  \P^3 & \dashrightarrow & \P^7 \\
  (t:x:y:z) & \longmapsto & (t^3:t^2x:t^2y:t^2z:txy:tyz:tzx:xyz)
\end{array}\right.
\end{equation}
The nine quadrics that define $\H$ correspond to the relations
$(t^2x)(t^2y)=(t^3)(txy),$ $(t^2x)(tyz)=(t^3)(xyz),$ $(t^2x)(xyz)=(txy)(tzx)$ and their cyclic 
permutations under $x \longmapsto y \longmapsto z \longmapsto x.$ The full resolution, given in \cite{Rub02},
is
\[
0 \longrightarrow \mathcal{F}_7[-6] \longrightarrow \mathcal{F}_7[-4]^9  \longrightarrow \mathcal{F}_7[-3]^{16} \longrightarrow \mathcal{F}_7[-2]^9 \longrightarrow \mathcal{F}_7 \longrightarrow \mathcal{F}_7 / I_\H \longrightarrow 0.
\]  

\subsection{The twisted case}

We will first define the twisted variety $\E$ of $\H$, for $\H$ the Segre embedding of $d$ copies of $\P^1$. We will then show how similar to the case $d=2$
the $q^d+1$ rational points $\E(\F_q)$ are in bijection with the $q^d+1$ rational points $\P^1(\F_{q^d})$. Finally this allows us to interpret the evaluation codes
$C_\E(s)$ as extended BCH codes. 
Set $r:=2^d-1$. For $d \geq 2$, let $\phi : \P^d \dashrightarrow \P^r$ be the natural rational map with image in $\H \subset \P^r$. The special case $d=3$ 
is given by (\ref{EmbeddingHyper2}). 

\begin{defn} \label{twistd}
For $d \geq 2$, let $(x_0:x_1:\cdots:x_d)$ be coordinates for $\P^d$, let $\alpha_1, \ldots, \alpha_d$ be an $\F_q$--basis of $\F_{q^d}$ and let
\[
\lambda : \left\{\begin{array}{ccccc}
  \P^d  & {\longrightarrow} & \P^d \\
  (x'_0:x'_1:\ldots:x'_d) & \longmapsto     & (x_0:x_1:\ldots:x_d)
\end{array}\right. 
\]
be the $\F_{q^d}$-linear transformation
$$
\left\{
  \begin{array}{ccl}
  x_0 & := & x'_0 \\
  x_j & := & \alpha_1^{q^{j-1}} x'_1 +\cdots + \alpha_d^{q^{j-1}} x'_d, \quad \textrm{for}\ j \in \{1, \ldots , d\}   
  \end{array} \right. .
$$
The rational map $\phi \circ \lambda : \P^d \dashrightarrow \P^r$ factors as $\mu_{tw} \circ \phi' : \P^d \dashrightarrow \P^r$ for a linear transformation 
$\mu_{tw} : \P^r \longrightarrow \P^r$ over $\F_{q^d}$ and a rational map $\phi' : \P^d \dashrightarrow \P^r$ over $\F_q$. The embedding $\phi'$ is called the twisted embedding with image $\E$.
\end{defn}

We illustrate the twisted embeddings for the cases $d=2$ and $d=3$.

\begin{exmp}
For $d=2$, the variety $\H \subset \P^3$. Over $\F_q$ it contains the rational points $(1:x:y:xy)$, for $x,y \in \F_q$. 
For the twisted variety $\E$, let $\{ b, b^q \}$ be a basis for $\F_{q^2} / \F_q$ and let 
\[
\left( \begin{array}{c} x \\ y \end{array} \right) = A \left( \begin{array}{c} u \\ v \end{array} \right), \qquad \text{for }
A = \left( \begin{array}{ll} b &b^q \\ b^q &b  \end{array} \right).
\]
Then
\[
(1:x:y:xy)^T = (I_1 \oplus A \oplus I_1) (1:u:v:Q)^T,
\]
for $xy = (bu+b^qv)(b^qu+bv) =: Q(u,v)$ irreducible of degree two over $\F_q$. The rational map $\phi'$ is given by
 \begin{equation}\label{EmbeddingEllBis}
\phi' : \left\{\begin{array}{ccc}
  \P^2 & \dashrightarrow & \P^3 \\
  (1:u:v) & \longmapsto & (1:u:v:Q)
\end{array}\right. .
\end{equation}
The finite rational points on the image $\E$ correspond to $\{ (1:u:v:Q) : u,v \in \F_q \}$. 
\end{exmp}

\begin{exmp}
For $d=3$, the variety $\H \subset \P^7$. Over $\F_q$ it contains the rational points $(1:x:y:z:xy:yz:zx:xyz)$ for $x,y,z \in \F_q$. 
For the twisted variety $\E$, let $\{ c, c^q, c^{q^2} \}$ be a basis for $\F_{q^3} / \F_q$ and let 
\[
\left( \begin{array}{c} x \\ y \\ z \end{array} \right) = A \left( \begin{array}{c} u \\ v \\ w \end{array} \right), \qquad \text{for }
A = \left( \begin{array}{lll} c &c^q &c^{q^2} \\ c^q &c^{q^2} &c  \\ c^{q^2} &c &c^q  \end{array} \right).
\]
Then 
\begin{multline}
(1:x:y:z:xy:yz:zx:xyz)^T  = (I_1 \oplus A \oplus B \oplus I_1)(1:u:v:w:Q_1:Q_2:Q_3:R)^T,
\end{multline}
for $xyz =: R(u,v,w)$ irreducible of degree three over $\F_q$, and for a $\F_{q^d}$-linear transformation $B$ and polynomials $Q_1, Q_2, Q_3$ of degree two over $\F_q$. The rational map $\phi'$ is given by
 \begin{equation}\label{EmbeddingHyperBis}
\phi' : \left\{\begin{array}{ccc}
  \P^3 & \dashrightarrow & \P^7 \\
  (1:u:v:w) & \longmapsto & (1:u:v:w:Q_1:Q_2:Q_3:R)
\end{array}\right. .
\end{equation}
The finite rational points on the image $\E$ correspond to $\{ (1:u:v:w:Q_1:Q_2:Q_3:R) : u,v,w \in \F_q \}$. 
A convenient choice for the polynomials $Q_1, Q_2, Q_3$ is as partial derivatives of the polynomial $R(u,v,w)$. The partial derivatives of $R(u,v,w)$ are defined over $\F_q$ and up to a linear transformation over $\F_{q^3}$ correspond to the partial derivatives of $xyz$. We include the details.
\[
\left( \begin{array}{c} \pr / \pr_u \\ \pr / \pr_v \\ \pr / \pr_w \end{array} \right) 
= A^T \left( \begin{array}{c} \pr / \pr_x \\ \pr / \pr_y \\ \pr / \pr_z  \end{array} \right)
\]
In particular, for $xyz = R(u,v,w),$
\[
\left( \begin{array}{c} yz \\ zx \\ xy \end{array} \right) = \left( \begin{array}{c} \pr / \pr_x \\ \pr / \pr_y \\ \pr / \pr_z  \end{array} \right) (xyz)
= (A^T)^{-1} \left( \begin{array}{c} \pr / \pr_u \\ \pr / \pr_v \\ \pr / \pr_w \end{array} \right) R(u,v,w).
\]
\end{exmp} 

For the variety $\E$ defined over $\F_q$ we obtain evaluation codes $C_\E(s)$ defined over $\F_q$. To determine the parameters of the codes 
we use a bijection between the rational points $\E(\F_q)$ and the rational points $\P^1(\F_{q^d})$ of the projective line over $\F_{q^d}.$
In analogy with (\ref{embq1}) consider the $\F_q$--embedding
\begin{equation} \label{embqd}
\psi : \P^1  \overset{\phi_f}{\longrightarrow} \P^1 \times \P^1 \times \cdots \times \P^1 \overset{\phi_s}{\longrightarrow} \P^r,
\end{equation}
where 
\[
\phi_f : \left\{\begin{array}{ccccc}
  \P^1  & {\longrightarrow} & \P^1 \times \P^1 \times \cdots \times \P^1 \\
  (x:y) & \longmapsto     & ((x:y),(x^q:y^q),\ldots(x^{q^{d-1}}:y^{q^{d-1}})) 
\end{array}\right. ,
\]
and $\phi_s$ is the Segre embedding such that $\psi(x:y) = (y^{q^{d-1}+\cdots+q+1}: \cdots : x^{q^{d-1}+\cdots+q+1})$.
 
\begin{lem}\label{TheGoodMapd}
We have an $\F_{q^d}$--embedding of $\P^1$ 
\begin{equation}\label{Fqdemb} \psi_{tw} : = \mu^{-1}_{tw} \circ \psi: \P^1 \overset{\phi_f}{\longrightarrow} \P^1 \times \cdots \times \P^1 \overset{\phi_s}{\longrightarrow} \P^r \overset{\mu^{-1}_{tw}}{\longrightarrow} \P^r \end{equation}
 inducing a one-to-one map $\P^1(\F_{q^d}) \longrightarrow \E (\F_q)$.  
\end{lem}

\begin{proof}
The proof is similar to the proof of Lemma \ref{TheGoodMap}. Let $(x:1)$ be a finite rational point on the projective line over $\F_{q^d}$. 
If we write $(x:1)=(\alpha_1 x'_1 + \cdots + \alpha_d x'_d : 1)$, with $x'_1, \ldots, x'_d \in \F_q$, for $\alpha_1, \ldots, \alpha_d$ as in Definition \ref{twistd},
then the image of $(x:1)$ under $\psi$ differs from a finite rational point in $\E(\F_q)$ by the linear transformation $\mu_{tw}.$ 
\end{proof}

\begin{defn}\label{DefExtBCHd}
Consider the projective line $\P^1$ over $\F_{q^d}$ and let $0\leq s\leq q-1$ be an integer. For $m = s(q^d-1)/(q-1)$, denote by $B^{ext}(s)$ the subcode of 
$C_{\P^1}(m)$ spanned by the evaluation at $\P^1(\F_{q^d})$ of the forms
\begin{equation} \label{bchform}
\{ x^i y^{m-i} : \text{$i = i_0 + i_1 q + \cdots + i_{d-1} q^{d-1}$ and $0 \leq i_0, i_1, \ldots, i_{d-1} \leq s,$} \}.
\end{equation}
The code $B_0^{ext}(s)$ is defined as the subfield subcode $B^{ext}(s)_{|\F_q}$.
\end{defn}

The codes $B^{ext}(s)$ and $B_0^{ext}(s)$ admit the group $PSL(2,\F_{q^d})$ as a $3$-transitive automorphism group. After puncturing at
$(0:1)$ and $(1:0)$ the code $B_0^{ext}(s)$ is a BCH code of type $[q^d-1,(s+1)^d,q^d-1-m]$ (\cite[Proposition 12]{DuursmaPellikaan}).

\begin{thm}\label{ParamEll2}
  For all $s< q-1$, the code $C_\E (s)$ has parameters 
  $[q^d+1, (s+1)^d, q^d+1-s(q^d-1)/(q-1)]$. Moreover, the code $C_\E(s)$ is permutation equivalent with
  an extended BCH code.
\end{thm}

\begin{proof} 
The proof is similar to the proof of Theorem \ref{Main}. Because of the bijection induced by $\psi_{tw}$ in Lemma \ref{TheGoodMapd} between
the $\F_q$--rational points of $\E$ and the $\F_{q^d}$--rational points of $\P^1$, the code $C_\E(s) \otimes \F_{q^d}$ can be obtained by evaluating the pullbacks 
$\psi_{tw}^{\star} (\mathcal{F}_r(s )\otimes \F_{q^d})$ at the elements of $\P^1 (\F_{q^d})$. The linear transformation $\mu_{tw}$ does not affect the code
over $\F_{q^d}$ and it suffices to consider the pullbacks $\psi^{\star} (\mathcal{F}_r(s )\otimes \F_{q^d})$. 
The definition of $\psi$ in (\ref{embqd}) entails that  $\psi^{\star} (\mathcal{F}_r(s) \otimes \F_{q^d})$ is generated by forms $x^i y^{m-i}$ that, in affine form,
are the product of $s$ monomials chosen from
\begin{equation} \label{geomform}
\{ 1^{j_0} x^{j_1} (x^q)^{j_2} \cdots (x^{q^{d-1}})^{j_{d-1}} : 0 \leq j_0, j_1, \ldots, j_{d-1} \leq 1 \}.
\end{equation}
Every such product is of the form (\ref{bchform}). Conversely, each from in (\ref{bchform}) can be written as a product of $s$ monomials in (\ref{geomform}). The latter
is clear if for a given monomial $x^i y^{m-i}$ we use an ordering on $i_0, i_1, \ldots, i_{d-1}$ to choose the monomials needed for the product.    
Thus we have shown that $C_\E(s) \otimes \F_{q^d}$ is the code $B(s)$ in Definition \ref{DefExtBCHd}. This clearly implies that $C_\E(s)$ is permutation equivalent 
with the extended BCH code $B_0(s)$. Moreover, using the $3-$transitivity of the automorphism group it implies that the parameters of $C_\E(s)$ are as claimed (as in the proof of Proposition \ref{ParamExtBCH}). 
\end{proof}

    

We observe that the last theorem has applications in two directions. It shows first that the maximum
number of $\F_q$--rational zeros in $\E \subset \P^r$ of a homogeneous form of degree $s$ agrees with
the BCH bound, that is to say it can be obtained using fairly elementary coding theory and without
using geometric tools. On the other hand it gives certain BCH codes a geometric interpretation as 
evaluation codes on an algebraic variety.



\subsubsection*{Acknowledgements.}
The first author is supported by the French ANR Defis program under contract
ANR-08-EMER-003 (COCQ project).
 
\bibliographystyle{abbrv}
\bibliography{biblio}

\bigskip

\noindent Alain Couvreur\\
Institut de Math\'ematiques de Bordeaux\\
Universit\'e Bordeaux I\\
351, cours de la Lib\'eration\\
33405 Talence Cedex, France\\
\url{couvreur@math.u-bordeaux1.fr}

\bigbreak

\noindent Iwan Duursma \\
Department of Mathematics\\
University of Illinois at Urbana--Champaign\\
1409 W. Green Street (MC-382) \\
Urbana, Illinois 61801-2975\\
\url{duursma@math.uiuc.edu}

\end{document}